\documentclass[amsmath, amssymb, 10pt]{iopart}
\usepackage{graphicx}
\begin{document}
\title{Linear aspects of the KKR formalism}
\author{T. Stopa, S. Kaprzyk and J. Tobo\l{}a}
\address{Faculty of Physics and Nuclear Techniques, \\
AGH University of Science and Technology, 30-059 Krak\'ow, Poland}
\date{\today}
\begin{abstract}
	We present one-dimensional KKR method with the aim to elucidate 
its linear features, particularly important in optimizing the numerical 
algorithms in energy bands computations. The conventional KKR equations 
based on the multiple scattering theory as well as novel forms of 
the secular matrix with nearly linear energy dependency of the 
eigenvalues are presented. The quasi-linear behaviour of these eigenvalue 
functions appears after (i) re-normalizing the wave functions in such a 
way that 'irregular' solutions vanish on the boundary of the 'muffin-tin' 
segments and (ii) integrating the full Green function over the whole
Wigner-Seitz cell. In addition, using the aforementioned approach we 
derive one-dimensional analog of the generalized Lloyd formula.

The novel KKR approach illustrated in one-dimension can be almost
directly applied to the higher dimensional cases. This should open 
prospects for the accurate KKR band structure computations of very 
complex materials.
\end{abstract}
\section{Introduction}
In the Korringa-Kohn-Rostoker (KKR) method the main problem in performing
electronic structure calculations is due to strongly nonlinear 
dependence of the \mbox{KKR-matrix} eigenvalues on energy for a 
fixed $k$-point. It is very difficult to find properly all the
zeros of eigenvalue functions, when there is a large number of atoms 
in the unit cell. Hence the fast and accurate extraction of all
zeros of eigenvalues still remains a challenge in computation of complex
materials.

Butler \cite{Butler} and Schwitalla \begin{it}et al.\end{it} \cite{Gyorffy} 
formulated the KKR formalism for one spatial dimension, which is very 
similar to its three-dimensional counterpart. The 
one-dimensional case is computationally much simpler and allows to 
formulate the substantial results in compact, analytical forms. 
Therefore, it is very attractive to investigate future of the KKR 
method along this line.

In this paper we use the Green function approach for solving the Schr\"odinger
equation. As already known, the one-dimensional (1D) KKR method gives exact 
solutions. This is due to the facts that (i) the unit cell can be entirely 
filled by the non-overlapping (touching each other) 'muffin-tin' segments and 
(ii) the multipole expansion contains only two 'spherical harmonics' 
($l=0$ and $l=1$). Furthermore, the KKR structure constants in the 1D case 
can be obtained analytically, which markedly facilitates numerical 
calculations.

Our paper is organized as follows. In Sec. 2 we recall after Butler
\cite{Butler} and Schwitalla \begin{it}et al.\end{it} \cite{Gyorffy} the
standard 1D-KKR formalism. In this section we also discuss few difficulties 
in determining the band structure in multi-atom systems if using the standard 
approach. Sec. 3 presents the novel form of the KKR-matrix in terms of 
logarithmic derivatives as well as the advantages of this approach with 
respect to the standard KKR theory. Next, the generalized Lloyd 
formula \cite{Kaprzyk} is derived , which helps in computation of the total 
number of states and placing properly the Fermi energy at early stage 
of calculations in real materials. In Sec. 5 we show illustrative 
results for an arbitrary chosen potential, which are followed by summary
and conclusions (Sec. 6).

\section{Standard KKR formalism in one dimension}
We look for a solution of the 1D Schr\"odinger equation with
Hamiltonian (atomic units are used, energy unit is 1 Ry):
\begin{equation}
\label{kkr1}
H=-\frac{d^2}{dx^2}+V(x).
\end{equation}
The potential in (\ref{kkr1}) is assumed to be translationally invariant with
the period of lattice constant $a$, so it can be written as
\begin{equation}
\label{kkr2}
V(x)=\sum_{n=-\infty}^{+\infty}v(x-na).
\end{equation}
We assume $v(x)$ in (\ref{kkr2}) to vanish for $|x|$ greater than some radius
$S$, and do not overlap each other. This assumption has the form analogous to
the non-overlapping 'muffin-tin' potential in three-dimensional (3D) KKR theory. 
We shall expand the potential $v(x)$ in (\ref{kkr2}) and the wave functions around
each center into symmetric and antisymmetric functions, analogous to the 3D
expansion into the spherical harmonics. First, let us define 1D spherical
coordinates centered at $na$-point,
\begin{equation}
\label{kkr3}
x=\hat{x}r+na, \hspace{1cm} x\in\left[na-\frac{a}{2}, na+\frac{a}{2}\right],
\end{equation}
with discrete directional coordinate, $\hat{x}=$sgn$(x)$,
and radius $r\in\left[0,\frac{a}{2}\right]$.
Then, with the definition of 1D analog of the spherical harmonics
\begin{equation}
\label{kkr4}
Y_0(\hat{x})=1/\sqrt{2} \hspace{1cm} Y_1(\hat{x})=\hat{x}/\sqrt{2},
\end{equation}
we decompose $v(x)$ into spherical and aspherical parts:
\begin{equation}
\label{kkr5}
v(x)=\frac{1}{2}[v(r)+v(-r)]+\frac{\hat{x}}{2}[v(r)-v(-r)]=
v_0(r)Y_0(\hat{x})+v_1(r)Y_1(\hat{x}).
\end{equation}
Similarly, we may write 1D multipole series for the wave function
\begin{equation}
\label{kkr6}
\psi(x)=\psi_0(r)Y_0(\hat{x})+\psi_1(r)Y_1(\hat{x})
\end{equation}
with the radial functions of the free Schr\"odinger equation given by the
following regular
\begin{equation}
\label{kkr7}
j_0(\sqrt{E}r)=\cos(\sqrt{E}r) \hspace{1cm} j_1(\sqrt{E}r)=\sin(\sqrt{E}r)
\end{equation}
and 'irregular'
\begin{equation}
\label{kkr8}
n_0(\sqrt{E}r)=\sin(\sqrt{E}r) \hspace{1cm} n_1(\sqrt{E}r)=-\cos(\sqrt{E}r)\;
\end{equation}
special solutions. Also analogs of the 3D spherical Hankel functions, usually
defined as $h_l(x)=j_l(x)+in_l(x)$, exist:
\begin{equation}
\label{kkr9}
h_0(\sqrt{E}r)=\exp(i\sqrt{E}r) \hspace{1cm} h_1(\sqrt{E}r)=-i\exp(i\sqrt{E}r)\;.
\end{equation}
The square root $\sqrt{E}$ is taken on the complex energy plane with the
branch cut along the positive real axis. To proceed further we follow 
the steps in the multiple scattering formalism \cite{Faulkner,Mills},
leading to the following form of the Bloch-Fourier Green's function
\begin{equation}
\label{kkr10}
\begin{array}{l}
<x'|G(E,k)|x>=-\sum_lJ_l(E,x_>)Z_l(E,x_<)\\
\mbox{}\\
+\sum_{l',l}Z_{l'}(E,x')\left[t^{-1}(E)-B(E,k)\right]^{-1}_{l'l}Z_l(E,x)~.
\end{array}
\end{equation}
In (\ref{kkr10}) $Z_l(E,x)$ and $J_l(E,x)$ are the regular and irregular solutions
of the Schr\"odinger equation with the radial parts given in the asymptotic
region $(r\geq S)$ by
\begin{equation}
\label{kkr11}
[Z(E,r)]_{l'l}=j_{l'}(\sqrt{E}r)[t^{-1}(E)]_{l'l}-\frac{i}{\sqrt{E}}h_l(\sqrt{E}
r ) \delta_{l'l}
\end{equation}
\begin{equation}
\label{kkr12}
[J(E,r)]_{l'l}=j_{l'}(\sqrt{E}r)\delta_{l'l}\;.
\end{equation}
$B(E,k)$ is the matrix of KKR structure constants with explicit form
generalized to multi-atom case as given in Appendix. The matrix $t(E)$ 
is one-scatterer $t$-matrix and may be found from the wave functions
\begin{equation}
\label{kkr13}
\psi_l(E,x)=\sum_{l'}Y_{l'}(\hat{x})[\psi(E,r)]_{l'l}
\end{equation}
with radial solutions satisfying the following set of differential equations
\begin{equation}
\label{kkr14}
\frac{d^2}{dr^2}\psi(E,r)+[E-v(r)]\psi(E,r)=0,
\end{equation}
with the potential $v(r)$ in the matrix form
\begin{equation}
\label{kkr15}
v(r)=\left[
\begin{array}{cc}
v_s(r) ,&v_a(r)\\
v_a(r) ,&v_s(r)
\end{array}
\right]=\frac{1}{\sqrt{2}}\left[
\begin{array}{cc}
v_0(r) ,&v_1(r)\\
v_1(r) ,&v_0(r)
\end{array}\right]\;.
\end{equation}
The functions $\psi(E,r)$ are routinely available in computations starting
at near origin ($r=0$) with
\begin{equation}
\label{kkr16}
[\psi(E,r)]_{l'l}=r^l\delta_{l'l}\;.
\end{equation}
If in asymptotic region, for $r\geq S$, these radial functions behave as
\begin{equation}
\label{kkr17}
[\psi(E,r)]_{l'l}=j_{l'}(\sqrt{E}r)[C(E)]_{l'l}-\frac{i}{\sqrt{E}}
h_{l'}(\sqrt{E}r)[S(E)]_{l'l}
\end{equation}
then
\begin{equation}
\label{kkr18}
[C(E)]_{l'l}=W\left\{[\psi(E,r)]_{l'l},
-\frac{i}{\sqrt{E}}h_{l'}(\sqrt{E}r)\right\}_{r=S},
\end{equation}
\begin{equation}
\label{kkr19}
[S(E)]_{l'l}=W\left\{j_{l'}(\sqrt{E}r), [\psi(E,r)]_{l'l}\right\}_{r=S}
\end{equation}
with the Wronskian definition
\begin{equation}
\label{kkr20}
W\{f,g\}=f\frac{\partial}{\partial r}g-g\frac{\partial}{\partial r}f.
\end{equation}
Comparing (\ref{kkr11}) and (\ref{kkr17}) gives us the following relation 
for the $t$-matrix
\begin{equation}
\label{kkr21}
t(E)=S(E)C^{-1}(E)\;.
\end{equation}
The poles of GF in (\ref{kkr10}) determine energy dispersion curves in the
form of bands. These poles are usually found from zeros of the
KKR-determinant
\begin{equation}
\label{kkr22}
\det\left|t^{-1}(E)-B(E,k)\right|=0.
\end{equation}
Furthermore, we define the matrix of logarithmic derivatives
\begin{equation}
\label{kkr23}
D(E,r)=\frac{\partial}{\partial r}[\psi(E,r)][\psi(E,r)]^{-1}.
\end{equation}
The $D(E,r)$ matrix is not affected by changing of the 
regular solution normalization. Moreover, this matrix is 
symmetrical, i.e.
\begin{equation}
\label{kkr24}
D^{T}(E,r)=D(E,r)
\end{equation}
what can be proved by recalling that it satisfies equation of 
Ricatti type
\begin{equation}
\label{kkr25}
\frac{\partial}{\partial r}D(E,r)+D^2(E,r)=v(r)-E
\end{equation}
with the symmetrical potential matrix, $v^{T}(r)=v(r)$. At the 
'muffin-tin' boundary point ($r=S$), we can simply write
\begin{equation}
\label{kkr26}
D(E)=\left.\frac{\partial}{\partial r}[\psi(E,r)]\right|_{r=S}[\psi(E,S)]^{-1}.
\end{equation}
The log-derivative matrix $D(E)$ is also directly related to 
the $t$-matrix by the following expression:
\begin{equation}
\label{kkr27}
\begin{array}{ll}
[v^{-1}_P(E)]_{l'l}&=j_{l'}(\sqrt{E}S)[t^{-1}(E)]_{l'l}j_l(\sqrt{E}S)\\
&\mbox{}\\
&-\displaystyle\frac{i}{\sqrt{E}}h_{l'}(\sqrt{E}S)j_l(\sqrt{E}S)\delta_{l'l}
\end{array}
\end{equation}
with the definition of the pseudopotential amplitude matrix
\begin{equation}
\label{kkr28}
v^{-1}_P(E)=D(E)-D_0(E)
\end{equation}
where
\begin{equation}
\begin{array}{l}
\label{kkr29}
D_0(E)=\displaystyle\frac{d}{dr}\left.[j(\sqrt{E}r)]\right|_{r=S}[j(\sqrt{E}S)]^{-1},\\
\mbox{}\\
{}[j(\sqrt{E}r)]_{l'l}=j_l(\sqrt{E}r)\delta_{l'l}\;.
\end{array}
\end{equation}

\section{Novel forms of the KKR-matrix in one dimension}
In the standard KKR-methodology finding zeros of the matrix $[t^{-1}(E)-B(E,k)]$ in
the GF formula (\ref{kkr10}) is central topic when computing energy bands. 
This is done by the requirement that at least one of the eigenvalues of that 
matrix goes through zero. In fact that condition comes out from analyzing 
not the full GF in (\ref{kkr10}) but only from its second part. 
To circumvent this problem, we want the first term in Eq.(\ref{kkr10}) 
to vanish at the boundary radial points. It can be done by re-normalizing 
the regular solution as
\begin{equation}
\label{n1}
\xi(E,r)=Z(E,r)Z^{-1}(E,S)
\end{equation}
and accordingly the irregular solution
\begin{equation}
\label{n2}
\zeta(E,r)=J(E,r)Z^T(E,S)-Z(E,r)j(\sqrt{E}S).
\end{equation}
By direct calculations we may check that the Wronskian in 
matrix form defined as
\begin{equation}
\label{n3}
W\left\{\zeta(E,r), \xi(E,r)\right\}_{l'l}=
\left\{\zeta^T(E,r)\frac{\partial}{\partial r}\xi(E,r)-
\frac{\partial}{\partial r}\zeta^T(E,r)\xi(E,r)\right\}_{l'l}
\end{equation}
is equal to
\begin{equation}
\label{n4}
W\{\zeta(E,r), \xi(E,r)\}=W\{J(E,r), Z(E,r)\}=\bf{1}.
\end{equation}
Setting in (\ref{n1}) and (\ref{n2}) $r=S$ and recalling 
that $t^T(E)=t(E)$ it follows, that
\begin{equation}
\label{n5}
\xi(E,S)=\begin{bf}1\end{bf} \hspace{1cm} \zeta(E,S)=0.
\end{equation}
Now, we are in position to convert GF expressed in (\ref{kkr10}) 
using convention $(J,Z)$ to that in terms of $(\zeta, \xi)$. 
First, we need the free GF matrix with radii on 'muffin-tin' 
points:
\begin{equation}
\label{n6}
\begin{array}{ll}
[g_0(E,k)]_{l'l}&=-\displaystyle\frac{i}{\sqrt{E}}h_l(\sqrt{E}S)j_l(\sqrt{E}S)\delta_{l'l}\\
&\mbox{}\\
&+j_{l'}(\sqrt{E}S)[B(E,k)]_{l'l}j_l(\sqrt{E}S)\;.
\end{array}
\end{equation}
Then applying the operator identity
\begin{equation}
\label{n7}
[A-B]^{-1}=A^{-1}+A^{-1}\left[B^{-1}-A^{-1}\right]^{-1}A^{-1}
\end{equation}
to the second term in (\ref{kkr10}) and replacing radial 
functions $(J,Z)$ with functions $(\zeta, \xi)$ as introduced 
in (\ref{n1}) and (\ref{n2}) we get to the following expression of GF: 
\begin{equation}
\label{n8}
<x'|G(E,k)|x>=\sum_{l_2l_1}Y_{l_2}(\hat{x}')<l_2\;r'|G(E,k)|l_1\;r>Y_{l_1}(\hat{x})
\end{equation}
with radial parts
\begin{equation}
\label{n9}
\begin{array}{l}
<l_2\;r'|G(E,k)|l_1\;r>=-\sum\limits_{l}[\zeta(E, r_>)]_{l_2l}[\xi(E, r_<)]^T_{ll_1}\\
\mbox{}\\
+\sum\limits_{l'l}[\xi(E, r')]_{l_2l'}\left[g_0^{-1}(E,k)-v_P(E)\right]_{l'l}^{-1}
[\xi(E,r)]^T_{ll_1}\;.
\end{array}
\end{equation}
It follows from (\ref{n9}) that GF with radial arguments at boundary points
\begin{equation}
\label{n10}
[g(E,k)]_{l_2l_1}\equiv <l_2\;S|G(E,k)|l_1\;S>
\end{equation}
can be found from algebraic equation of Dyson type
\begin{equation}
\label{n11}
g(E,k)=g_0(E,k)+g_0(E,k)v_P(E)g(E,k)
\end{equation}
with pseudopotential amplitude $v_P(E)$ as given in (\ref{kkr28}). 
The poles of $g(E,k)$ are exactly the same as of the full GF 
$<x'|G(E,k)|x>$ in (\ref{kkr10}), suggesting alternative way 
to (\ref{kkr22}) of finding energy bands
\begin{equation}
\label{n12}
\det|g^{-1}(E,k)|=0.
\end{equation}
Searching for zeros of eigenvalues of matrix $g^{-1}(E,k)$ is much easier than 
that of \mbox{$t^{-1}(E)-B(E,k)$} as the eigenvalue functions of the former are 
monotonically increasing with energy $E$. But some obscuring deficiency 
still persists as the slopes, at which these eigenvalues are crossing 
energy axis, are not fixed. This inconvenience can be avoided in the 
following way. Let us first integrate GF in the form (\ref{n9}) over 
the whole \mbox{Wigner-Seitz} cell
\begin{equation}
\label{n13}
\begin{array}{l}
\int\limits_{-a/2}^{a/2}dx<x|G(E,k)|x>=-\sum\limits_l\int\limits_{-a/2}^{a/2}dx\;
\zeta_l(E,x)\xi_l(E,x)
\\ \mbox{}\\
+\sum\limits_{l'l}[g(E,k)]_{l'l}\int\limits_{-a/2}^{a/2}dx\;\xi_{l'}(E,x)\xi_l(E,x).
\end{array}
\end{equation}
Integrating the first term in (\ref{n13}) yields:
\begin{equation}
\label{n14}
\begin{array}{l}
\int\limits_{-a/2}^{a/2}dx\zeta_{l'}(E,x)\xi_l(E,x)=\\
\mbox{}\\
\sum_{l_2l_1}\int\limits_0^Sdr\sum_{\hat{x}}Y_{l_2}(\hat{x})Y_{l_1}(\hat{x})
[\zeta(E,r)]_{l_2l'}[\xi(E,r)]_{l_1l}=\\
\mbox{}\\
\int\limits_0^Sdr[\zeta^T(E,r)\xi(E,r)]_{l'l}\\
\end{array}
\end{equation}
From the Schr\"odinger equation we find that radial parts $\zeta^T(E,r)$ 
and the energy derivative $\frac{\partial}{\partial E}
\xi(E,r)\equiv\dot{\xi}(E,r)$ satisfy differential equations
\begin{equation}
\label{n15}
\frac{d^2}{dr^2}\zeta^T(E,r)=\zeta^T(E,r)[v(r)-E]
\end{equation}
and
\begin{equation}
\label{n16}
\frac{\partial^2}{\partial r^2}\dot{\xi}(E,r)=[v(r)-E]\dot{\xi}(E,r)-\xi(E,r)
\end{equation}
respectively. Multiplying (\ref{n15}) to the right by $\xi(E,r)$, 
then (\ref{n16}) to the left by $\zeta^T(E,r)$ and subtracting one 
from another, we get: 
\begin{equation}
\label{n17}
\int\limits_0^Sdr\zeta^T(E,r)\xi(E,r)=
W\left\{\zeta(E,r),\dot{\xi}(E,r)\right\}_{r=0}
\end{equation}
This integral can be conventionally found if function $\psi(E,r)$ (\ref{kkr16}) 
is computed. From the definition of $\xi(E,r)$ 
in (\ref{n1}) it follows:
\begin{equation}
\label{n18}
\xi(E,r)\psi(E,S)=\psi(E,r),
\end{equation}
\begin{equation}
\label{n19}
\dot{\xi}(E,r)\psi(E,S)+\xi(E,r)\dot{\psi}(E,S)=\dot{\psi}(E,r)=0
\hspace{0.5cm}(r\rightarrow 0),
\end{equation}
Inserting $\dot{\xi}(E,r)$ from (\ref{n19}) into (\ref{n17}) we get 
\begin{equation}
\label{n20}
\begin{array}{l}
W\left\{\zeta(E,r),\dot{\xi}(E,r)\right\}_{r=0}=\\
\mbox{}\\
W\left\{\zeta(E,r),-\xi(E,r)\dot{\psi}(E,S)\psi^{-1}(E, S)\right\}_{r=0}=\\
\mbox{}\\
-\dot{\psi}(E,S)\psi^{-1}(E,S)
\end{array}
\end{equation}
Combining (\ref{n17}) with the result (\ref{n20}) we have
\begin{equation}
\label{n21}
\int_0^Sdr\zeta^T(E,r)\xi(E,r)=-\dot{\psi}(E,S)\psi^{-1}(E,S)\;.
\end{equation}
For the second integral in (\ref{n13}) we obtain
\begin{equation}
\label{n22}
\begin{array}{l}
\int\limits_{-a/2}^{a/2}dx\;\xi_{l'}(E,x)\xi_l(E,x)=\int\limits_0^Sdr
[\xi^T(E,r)\xi(E,r)]_{l'l}=\\
\mbox{}\\
{[}N_{\psi}(E)\psi^{-1}(E,S){]}^T{[}N_{\psi}(E)\psi^{-1}(E,S){]}
\end{array}
\end{equation}
with the definition of $N_{\psi}(E)$ as
\begin{equation}
\label{n23}
\int\limits_0^Sdr\psi^T(E,r)\psi(E,r)=N^T_{\psi}(E)N_{\psi}(E)\;.
\end{equation}
With the help of (\ref{n21}) and (\ref{n22}) we may convert (\ref{n13}) 
to the form
\begin{equation}
\label{n24}
\begin{array}{l}
\int\limits_{-a/2}^{a/2}<x|G(E,k)|x>=\sum\limits_l[\dot{\psi}(E,S)\psi^{-1}(E,S)]_{l'l}+\\
\mbox{}\\
\sum\limits_l\left\{[N_{\psi}(E)\psi^{-1}(E,S)]g(E,k)[N_{\psi}(E)\psi^{-1}(E,S)]^T
\right\}_{ll}=\\
\mbox{}\\
\sum\limits_lP_{ll}(E,k)
\end{array}
\end{equation}
The left hand side of equation (\ref{n24}) is equal to the trace of 
hermitian matrix $P(E,k)$ defined as
\begin{equation}
\label{n25}
P(E,k)=1/2[Q(E,k)+Q^{\dagger}(E,k)],
\end{equation}
with
\begin{equation}
\label{n26}
\begin{array}{ll}
Q(E,k)=&[\psi(E,S)N_{\psi}^{-1}(E)]^{-1}[\dot{\psi}(E,S)N_{\psi}^{-1}(E)]+\\
&\mbox{}\\
&N_{\psi}(E)\psi^{-1}(E,S)g(E,k)[N_{\psi}(E)\psi^{-1}(E,S)]^T\;.
\end{array}
\end{equation}
The poles of the matrix $P(E,k)$ give energy bands that can be 
found from zeros of the determinant
\begin{equation}
\label{n27}
\det|P^{-1}(E,k)|=0\;.
\end{equation}
From the analytical property of GF known in the literature as Herglotz 
property \cite{Ratana} it is expected that eigenvalues of $P^{-1}(E,k)$-matrix 
must increase monotonically with energy, i.e.,
\begin{equation}
\label{n28}
\frac{\partial}{\partial E}\lambda_i[P^{-1}(E,k)]\geq 0,
\end{equation}
and with the slope equal to one if crossing energy axis.

In the formula (\ref{n26}) the presence of the inverse matrix 
$\psi^{-1}(E,S)$ may cause numerical instability if matrix 
$\psi(E,S)$ is singular, what may happen in practice. 
To avoid such complications it is convenient to combine the
first and second term in Eq.(\ref{n25}) together with the 
following manipulation. First, we can compute integral in 
(\ref{n23}) using trick with energy derivative, like in 
(\ref{n15}) and (\ref{n16}). 
Then with the known procedure we get
\begin{equation}
\label{n31}
\int\limits_0^Sdr\psi^T(E,r)\psi(E,r)=
-W\left\{\psi(E,r),\dot{\psi}(E,r)\right\}_{r=S}.
\end{equation}
In equation (\ref{n31}) the contribution at the origin point $(r=0)$ is 
set to be equal zero, what is justified with normalization of 
$\psi(E,r)$ assumed in Eq.(\ref{kkr16}). Recalling Wronskian definition, 
as set in (\ref{kkr20}) and log-derivatives in (\ref{kkr26}), we may 
proceed with (\ref{n31}) getting
\begin{equation}
\label{n32}
\int_0^Sdr\psi^T(E,r)\psi(E,r)=\psi^T(E,S)\left\{D_{\psi}(E)-D_{\dot{\psi}}(E)
\right\}\dot{\psi}(E,S)
\end{equation}
From (\ref{n32}) and (\ref{n23}) it follows the equality
\begin{equation}
\label{n33}
[\psi(E,S)N^{-1}_{\psi}(E)]^T\left\{D_{\psi}(E)-D_{\dot{\psi}}(E)\right\}
[\dot{\psi}(E,S)N^{-1}_{\psi}(E)]=1,
\end{equation}
useful in converting (\ref{n26}) to the desired form as follows
\begin{equation}
\label{n34}
\begin{array}{l}
Q(E,k)=[\psi(E,S)N_{\psi}(E)]^{-1}\times\\
\mbox{}\\
\left\{\dot{\psi}(E,S)N_{\psi}^{-1}(E)+g(E,k)[\psi(E,S)N_{\psi}^{-1}(E)]
^{-1^T}\right\}
\end{array}
\end{equation}
From (\ref{n34}) with the help of equality (\ref{n33}) we may 
proceed to
\begin{equation}
\label{n35}
\begin{array}{l}
Q(E,k)=[\psi(E,S)N_{\psi}(E)]^{-1}\times\\
\mbox{}\\
\left\{[D_{\psi}(E)-D_{\dot{\psi}}(E)]^{-1}+g(E,k)\right\}
[\psi(E,S)N_{\psi}^{-1}(E)]^{-1^T}\;.
\end{array}
\end{equation}
But recalling $g(E,k)$ as given in (\ref{n11}) we get
\begin{equation}
\label{n36}
\begin{array}{l}
[D_{\psi}(E)-D_{\dot{\psi}}(E)]^{-1}+g(E,k)=\\
\mbox{}\\
{}[g_0^{-1}(E,k)+D_0(E)-D_{\psi}(E)]^{-1}
[g_0^{-1}(E,k)+D_0(E)-D_{\dot{\psi}}(E)]\times\\
\mbox{}\\
{}[D_{\psi}(E)-D_{\dot{\psi}}(E)]^{-1}.
\end{array}
\end{equation}
that if inserted in (\ref{n35}) gives
\begin{equation}
\label{n37}
\begin{array}{l}
Q(E,k)=\\
\mbox{}\\
\left\{\psi(E,S)N_{\psi}^{-1}(E)+g_0(E,k)[D_0(E)-D_{\psi}(E)]
\psi(E,S)N_{\psi}^{-1}(E)\right\}^{-1}\times\\
\mbox{}\\
\left\{\dot{\psi}(E,S)N_{\psi}^{-1}(E)+g_0(E,k)[D_0(E)-D_{\dot{\psi}}(E)]
\dot{\psi}(E,S)N_{\psi}^{-1}(E)\right\}\;.
\end{array}
\end{equation}
With the Wronskian definition as in (\ref{kkr20}) last expression 
can be simplified further
\begin{equation}
\label{n38}
\begin{array}{ll}
Q(E,k)=N_{\psi}(E)&\left\{\psi(E,S)+C(E,k)W^T[\psi,j]\right\}^{-1}\times\\
\mbox{}\\
&\left\{\dot{\psi}(E,S)+C(E,k)W^T[\dot{\psi},j]\right\}N_{\psi}^{-1}(E)
\end{array}
\end{equation}
with the matrix $C(E,k)$ formed from KKR structure functions $B(E,k)$:
\begin{equation}
\label{n39}
[C(E,k)]_{l'l}=-\frac{i}{\sqrt{E}}h_{l'}(\sqrt{E}S)\delta_{l'l}+
j_{l'}(\sqrt{E}S)[B(E,k)]_{l'l}
\end{equation}
These two last equations give the most convenient way to construct 
the novel \mbox{KKR-matrix} $P(E,k)$ with desired properties of the eigenvalues.

\section{Generalised Lloyd formula}
The Lloyd formula is often used to calculate the number of states for a
'muffin-tin' model potential. We derive it here in the form similar 
to that used in the paper by Kaprzyk and Bansil \cite{Kaprzyk}.

It is well known that the total number of states $N(E)$ is done as
\begin{equation}
\label{l1}
N(E)=-\frac{1}{\pi}{\textrm Im}\int_{-\infty}^EdE\sum_{k\in BZ}\int_{(WS)}dx
<x|G(E,k)|x>.
\end{equation}
We show that the space integral in (\ref{l1}) can be expressed as a perfect
energy derivative and the energy integration can be done explicitly. 
We start with (\ref{n24}) putting it in the form
\begin{equation}
\label{l2}
\begin{array}{l}
\int\limits_{-a/2}^{a/2}dx<x|G(E,k)|x>=\Tr\{\psi^{-1}(E,S)\dot{\psi}(E,S)\}+\\
\mbox{}\\
\Tr\{g(E,k)[N_{\psi}(E)\psi^{-1}(E,S)]^T[N_{\psi}(E)\psi^{-1}(E,S)]\}\;,
\end{array}
\end{equation}
then with the help of (\ref{n32}) we find (at $r=S$)
\begin{equation}
\label{l3}
N_{\psi}^T(E)N_{\psi}(E)=
-\psi^T(E,S)\dot{D}_{\psi}(E)\psi(E,S).
\end{equation}
With the last result, (\ref{l2}) can be written as
\begin{equation}
\label{l4}
\begin{array}{l}
\int\limits_{-a/2}^{a/2}dx<x|G(E,k)|x>=\\
\mbox{}\\
\Tr\left\{\psi^{-1}(E,S)\dot{\psi}(E,S)\right\}-
\Tr\left\{g(E,k)\dot{D}_{\psi}(E)\right\}.
\end{array}
\end{equation}
The last step consists of proving the equality
\begin{equation}
\label{l5}
\frac{\partial}{\partial E}g(E,k)=g(E,k)\dot{D}_{\psi}(E)g(E,k)
\end{equation}
which if inserted into (\ref{l4}) gives the result
\begin{equation}
\label{l6}
\int_{-a/2}^{a/2}dx<x|G(E,k)|x>=-\frac{\partial}{\partial E}
\Tr\ln[g(E,k)\psi^{-1}(E,S)].
\end{equation}
The algebraic equality (\ref{l5}) is a consequence of GF properties 
stating that
\begin{equation}
\label{l7}
\frac{\partial}{\partial E}G(E)=-G(E) G(E).
\end{equation}
If applied to our Bloch-Fourier GF with radial arguments on 'muffin-tin' 
boundary points we get
\begin{equation}
\label{l8}
\begin{array}{l}
<\hat{x}_2S|\frac{\partial}{\partial E}G(E,k)|\hat{x}_1S>=\\
\mbox{}\\
-\int\limits_{-a/2}^{a/2}dx<\hat{x}_2\;S|G(E,k)|x>
<x|G(E,k)|\hat{x}_1\;S>
\end{array}
\end{equation}
and for the radial part defined in (\ref{n8})
\begin{equation}
\label{l9}
\begin{array}{l}
<l_2S|\frac{\partial}{\partial E}G(E,k)|l_1S>=\\
\mbox{}\\
-\int\limits_0^Sdr<l_2\;S|G(E,k)|l\;r><l\;r|G(E,k)|l_1\;S>.
\end{array}
\end{equation}
Inserting proper radial arguments into (\ref{l9}) we get according 
to (\ref{n7}) and (\ref{n5}) that
\begin{equation}
\label{l10}
<l_2\;S|G(E,k)|l\;r>=[g(E,k)\zeta^T(E,r)]_{l_2l},
\end{equation}
\begin{equation}
\label{l11}
<l\;S|G(E,k)|l_1\;r>=[\xi(E,r)g(E,k)]_{ll_1},
\end{equation}
which allow us to rewrite (\ref{l9}) in the form
\begin{equation}
\label{l12}
\frac{\partial}{\partial E}\left[g(E,k)\right]=
-g(E,k)\left[\int\limits_0^Sdr\;\xi^T(E,r)\xi(E,r)\right]g(E,k).
\end{equation}
Computing integral in (\ref{l12}) can be done with the help of 
(\ref{l3}) and leads to the result
\begin{equation}
\label{l13}
\int\limits_0^Sdr\xi^T(E,r)\xi(E,r)=-\dot{D}_{\xi}(E)=-\dot{D}_{\psi}(E).
\end{equation}
Inserting this relation into (\ref{l12}) proves validity of (\ref{l5}). 
The total number of states seen in (\ref{l1}) can now be found from
\begin{equation}
\label{l14}
N(E)=\frac{1}{\pi}{\textrm Im}\ln(\det|\psi(E,S)|)
-\frac{1}{\pi}\sum_{k\in BZ}{\textrm Im}\ln\det|g(E,k)|.
\end{equation}
In passing from (\ref{l6}) to (\ref{l14}) we used algebraic equality
\begin{equation}
\label{l15}
\Tr\ln[A]=\ln[\det|A|].
\end{equation}
In (\ref{l14}) the first term accounts for the number of nodes on regular
solution from origin up to radius $S$, and the second term comes out from 
the Bloch states at each of $k$-points in BZ.

\section{Results}
\begin{figure}
\begin{center}
\includegraphics[angle=-90, scale=0.50]{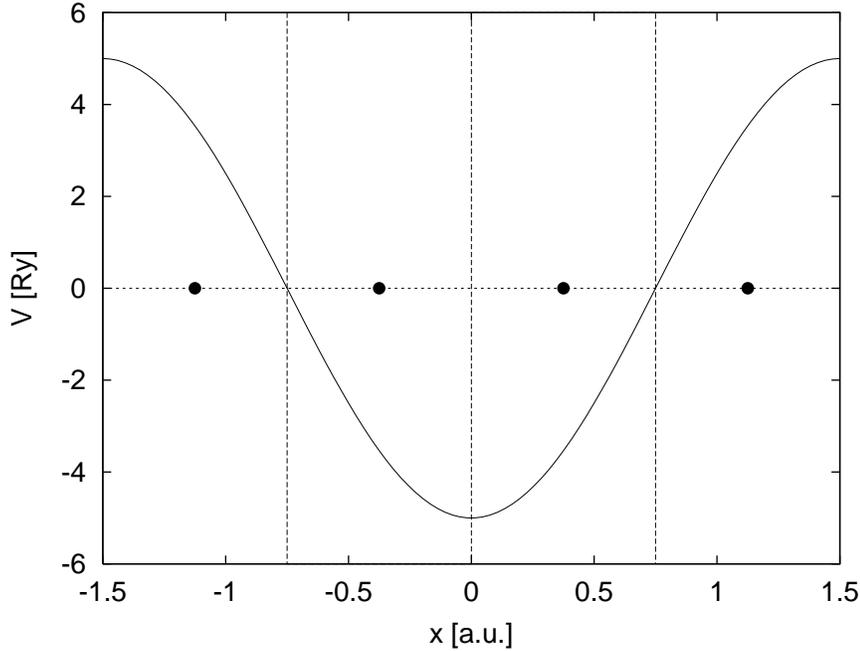}
\caption{\label{rys_pot}Mathieu potential used in calculations. 
The black dots denote four atoms in the unit cell case. In the case of 
one atom, it is placed in the middle of the unit cell.}
\end{center}
\end{figure}

\begin{figure}
\begin{center}
\includegraphics[angle=-90, scale=0.50]{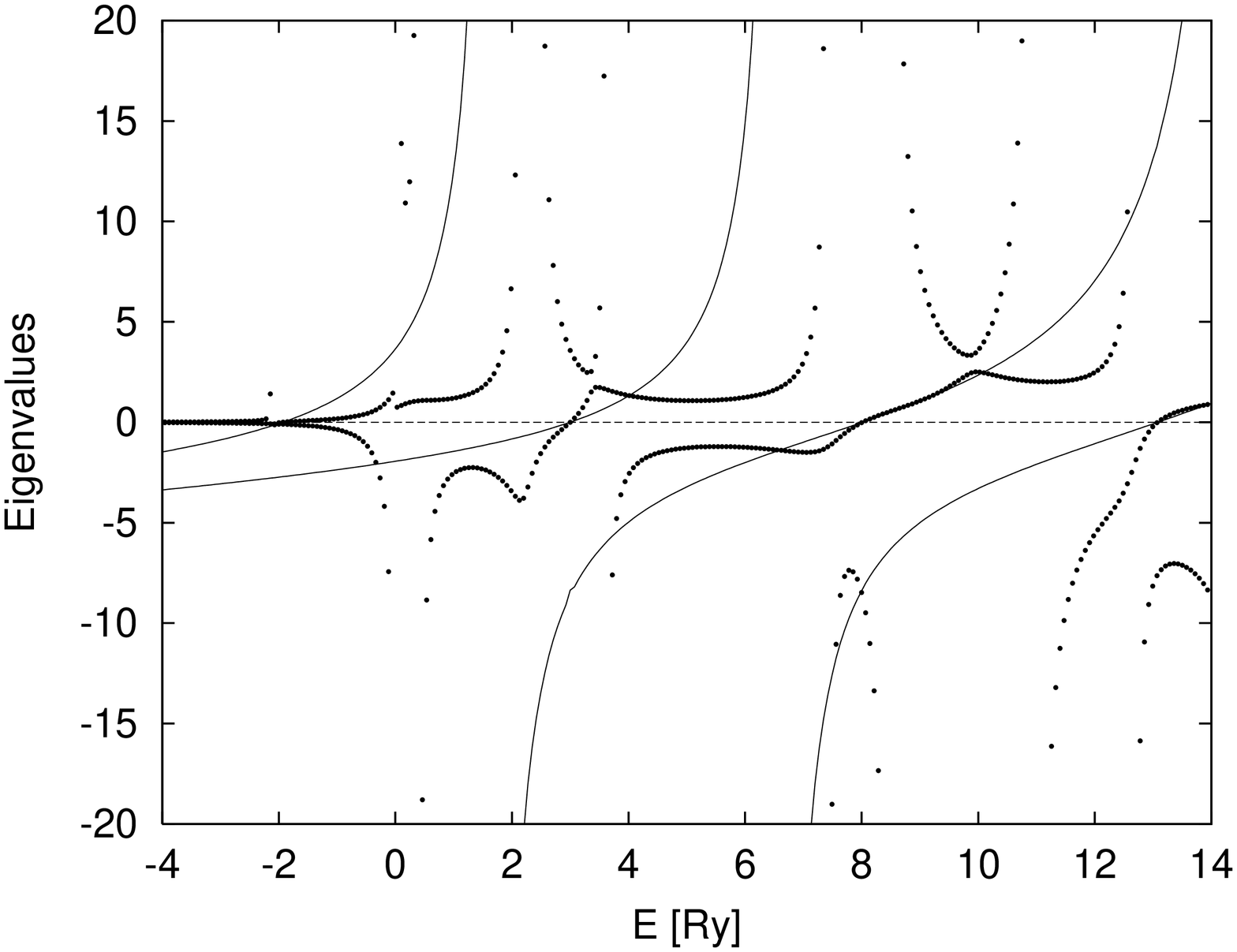}
\caption{\label{rys_comp}Comparison of the standard KKR method (dots) 
and the novel one (solid lines) as applied to the Mathieu potential (\ref{r1}). 
The wave vector is $k=0.6\pi/a$.}
\end{center}
\end{figure}

\begin{figure}
\begin{center}
\includegraphics[angle=-90, scale=0.50]{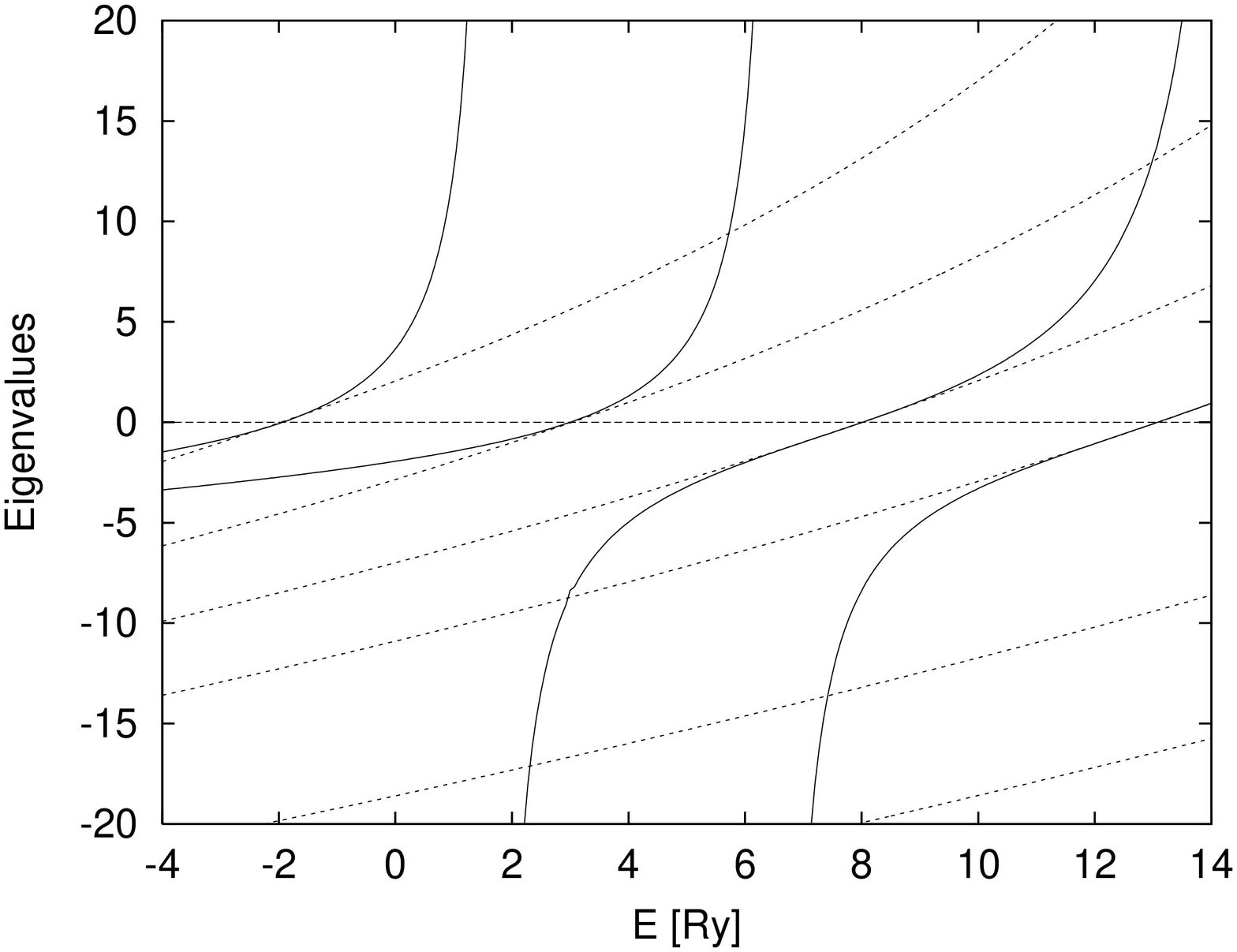}
\caption{\label{rys_eig}The eigenvalue functions for the Mathieu potential 
(\ref{r1}) and the wave vector $k=0.6\pi/a$. The case with one and four 
atoms in the unit cell is represented by solid and dashed lines, 
respectively.}
\end{center}
\end{figure}

In order to illustrate numerically how the novel approach to KKR method
works compared to standard one, we have performed calculations for 
the 1D Mathieu potential of the form
\begin{equation}
\label{r1}
v(x)=-U_0\cos(2\pi x/a),
\end{equation}
and used lattice constant $a=3.0$~a.u. and $U_0=5$~Ry (see Fig.~\ref{rys_pot}).
In Fig.~\ref{rys_comp} we show eigenvalue functions as calculated using 
both standard and novel method (Eqs.~(\ref{kkr22}) and (\ref{n27})) for 
the potential (\ref{r1}). One can see that zeros of the eigenvalue functions 
are exactly the same. In the novel method case we can find them numerically 
very easily but in the standard KKR approach it is much more difficult.
Moreover, to illustrate how the eigenvalue curves behave when the number 
of atoms in the unit cell grows, we performed the calculations in two cases. 
In the first case we put only one atom in the unit cell with the 'muffin-tin' 
radius $S=1.5$~a.u. and the symmetric potential given by (\ref{r1}). 
In the second case we divided the unit cell into four 'muffin-tin' segments 
of the same length and touching each other, but without changing the 
total potential in the unit cell. 
So, in this case the potential in each segment is not symmetrical, but the 
physical situation is exactly the same. What changed is the size of resulting 
matrices and hence number of the eigenvalue functions. These functions in both
cases are shown in Fig.~\ref{rys_eig} for the wave vector $k=0.6\pi/a$. 
Note that in the second case the lines are not derived from interpolation 
but simply connect calculated points, which are not shown in Fig.~\ref{rys_eig}. 

Now all the benefits of the novel method can be underlined. First, 
we observe that the eigenvalue functions do not cross each other and 
grow monotonically. Moreover, they form almost straight lines with 
the slope of unity when passing through zero. Secondly, we can see that 
with the increasing number of atoms in the unit cell the eigenvalue 
curves become straight for the wider range of energies. So, in the limit 
of infinite number of atoms in the unit cell, the method seems to 
become linear. This makes the method not very sensitive to the number 
of energy points used in the interpolation procedure, when finding zeros 
of eigenvalue curves. This last advantage opens possibilities of 
calculation of very complex systems, even with hundreds atoms in the 
unit cell in reasonable time without loosing accuracy.

\section{Summary}
In this paper we have reformulated one-dimensional full potential 
\mbox{KKR formalism}. Then we have derived novel form of the KKR 
secular matrix. To exclude first term in the full Green function 
we have normalized solutions of the Schr\"{o}dinger equation in 
such a way that the 'irregular' solution disappears at boundary points 
of 'muffin-tin' segments. Then by integrating the full Green function 
over the Wigner-Seitz cell we have derived the final result, that is 
the expressions for the secular matrix $P(E,k)$ as well as for the 
generalised Lloyd formula. The eigenvalues of the inverse matrix $P^{-1}(E,k)$ 
increase monotonically with energy and almost linearly for every $k$-point. 
In the case of increasing number of atoms in the unit cell the 
eigenvalue functions become more linear. 

Finally, we performed numerical calculations for the case of the Mathieu potential. 
The results show that the zeros of the eigenvalues can be easily found, even 
for very complex systems, without any lost of accuracy. This formalism can be 
extended to the higher-dimensional systems with the minor changes.
\section*{Appendix}
\appendix
\setcounter{section}{1}
In this appendix we briefly generalise the formalism introduced in sections
2, 3 and 4 for many atoms in the unit cell. In this case one has to
solve the Schr\"odinger equation with the potential
\begin{equation}
\label{a0}
V(x)=\sum_{n=-\infty}^{\infty}\sum_{\alpha=1}^{p}v_{\alpha}(x-na-a_{\alpha}),
\end{equation}
where $a_{\alpha}$ is the position of $\alpha$-th atom in the unit cell with $p$
equal to the number of atoms in the unit cell.
At each \mbox{'muffin-tin'} atomic segment we solve Schr\"odinger equation starting
at origin from
\begin{equation}
\label{a1}
[\psi^{\alpha}(E,r)]_{l'l}=r^l\delta_{l'l} \hspace{1cm} r\rightarrow 0\;.
\end{equation}
The resulting $\psi$ matrix has the form
\begin{equation}
\label{a2}
[\psi(E,r)]_{\alpha'l',\alpha l}=
[\psi^{\alpha}(E,r)]_{l'l}\delta_{\alpha\alpha'}
\end{equation}
with the size $2p\times 2p$. Similarly we construct all other matrices
introduced in the previous sections. GF in $(\zeta, \xi)$ representation now 
has a form:
\begin{equation}
\label{a21}
\begin{array}{l}
<x'+a_{\alpha'}|G(E,k)|x+a_{\alpha}>=\\
\mbox{}\\
-\sum\limits_{l}\zeta_l^{\alpha}(E,x_>)\xi_l^{\alpha}(E,x_<)\delta_{\alpha\alpha'}\\
\mbox{}\\
+\sum\limits_{l,l'}\xi_{l'}^{\alpha'}(E,x')
[g_0^{-1}(E,k)-v_P(E)]^{-1}_{\alpha'l', \alpha l}
\xi_{l}^{\alpha}(E, x)\;.
\end{array}
\end{equation}
The structure constants in this generalised case can be
given explicitly with the matrix elements $[B(E,k)]_{\alpha'l',\alpha l}$:
\begin{equation}
\label{a4}
\begin{array}{l}
[B(E,k)]_{\alpha 0,\alpha' 0}=[B(E,k)]_{\alpha 1,\alpha' 1}=\\
\mbox{}\\
\displaystyle\frac{\exp(i\sqrt{E} a)\cos\sqrt{E} a_{\alpha\alpha'}-\cos(ka-\sqrt{E}
 a_{\alpha\alpha'})}
{i\sqrt{E}[\cos ka-\cos\sqrt{E}a]}\\
\mbox{}\\
+\displaystyle\frac{1}{i\sqrt{E}}\exp(i\sqrt{E}|a_{\alpha\alpha'}|)
(1-\delta_{\alpha\alpha'}),
\end{array}
\end{equation}
\begin{equation}
\label{a5}
\begin{array}{l}
[B(E,k)]_{\alpha 1, \alpha' 0}=[B(E,k)]_{\alpha' 0, \alpha 1}^*=\\
\mbox{}\\
\displaystyle\frac{\exp(i\sqrt{E} a)\sin\sqrt{E} a_{\alpha\alpha'}+\sin(ka-\sqrt{E}
 a_{\alpha\alpha'})}{i\sqrt{E}[\cos ka-\cos\sqrt{E}a]}\\
\mbox{}\\
+\displaystyle\frac{1}{\sqrt{E}}\exp(i\sqrt{E}|a_{\alpha\alpha'}|)\textrm{sgn}(-a_{\alpha\alpha'})
(1-\delta_{\alpha\alpha'}),
\end{array}
\end{equation}
with $a_{\alpha\alpha'}=a_{\alpha}-a_{\alpha'}$.
Finally we define the matrix $P(E,k)$ by the relation:
\begin{equation}
\label{a6}
\sum_{\alpha}\int\limits_{-S_{\alpha}}^{S_{\alpha}}dx<x+a_{\alpha}|G(E,
 k)|x+a_{\alpha}>=
\Tr[P(E, k)]
\end{equation}
with $S_{\alpha}$ being muffin-tin radius of the $\alpha$-th atom.
Following the steps described in previous sections we find, that
the matrix $P(E,k)$ can be found using Eqs. (\ref{n25}) and (\ref{n26}) by
adding atomic indexes to the matrix elements, as it is done for structure
constants matrix $B(E,k)$. The band structure is still determined by the 
relation (\ref{n27}).


\end{document}